\documentclass[aps,prl,superscriptaddress,twocolumn]{revtex4-2}

\usepackage{amssymb}
\usepackage{graphicx}
\usepackage{dcolumn}
\usepackage{bm}
\usepackage{amsmath}
\usepackage[normalem]{ulem}
\usepackage{textcomp}
\usepackage{float}
\usepackage[usenames]{color}
\usepackage{physics}
\usepackage{blindtext}

\usepackage[unicode=true,bookmarks=true,bookmarksnumbered=false,bookmarksopen=false,breaklinks=false,pdfborder={0 0 1},backref=false,colorlinks=true]{hyperref}
\hypersetup{linkcolor=magenta,urlcolor=blue,citecolor=blue,pdfstartview={FitH},hyperfootnotes=false,unicode=true}

\def\be{\begin{equation}}
\def\ee{\end{equation}}
\def\bea{\begin{eqnarray}}
\def\eea{\end{eqnarray}}

\begin{document}

\title{Nagaoka supermetal in the particle-doped triangular Hubbard model}

\author{Rui Cao}
\affiliation{College of Science, National University of Defense Technology, Changsha 410073, P. R. China}
\author{Xiangyue Zhang}
\affiliation{College of Science, National University of Defense Technology, Changsha 410073, P. R. China}
\author{Hui Tan}
\affiliation{College of Science, National University of Defense Technology, Changsha 410073, P. R. China}

\author{Jian-Shu Xu}
\affiliation{Institute of Modern Physics, Northwest University, Xi'an 710127, P. R. China}

\author{Yuan-Yao He}
\email{heyuanyao@nwu.edu.cn}
\affiliation{Institute of Modern Physics, Northwest University, Xi'an 710127, P. R. China}
\affiliation{Shaanxi Key Laboratory for Theoretical Physics Frontiers, Xi'an 710127, P. R. China}
\affiliation{Fundamental Discipline Research Center for Quantum Science and Technology of Shaanxi Province, Xi'an 710127, P. R. China}
\affiliation{Hefei National Laboratory, Hefei 230088, P. R. China}

\author{Jianmin Yuan}
\affiliation{Institute of Atomic and Molecular Physics, Jilin University, Changchun 130012, P. R. China}
\affiliation{College of Science, National University of Defense Technology, Changsha 410073, P. R. China}

\author{Yongqiang Li}
\email{li\_yq@nudt.edu.cn}
\affiliation{College of Science, National University of Defense Technology, Changsha 410073, P. R. China}
\affiliation{Hunan Key Laboratory of Extreme Matter and Applications, National University of Defense Technology, Changsha 410073, P. R.  China}
\affiliation{Hunan Research Center of the Basic Discipline for Physical States, National University of Defense Technology, Changsha 410073,  P. R. China}

\begin{abstract}
While the interplay of correlations and geometric frustration in doped Mott insulators provides a fertile ground for exotic quantum phases, the nature of the metallic state emerging upon particle doping remains poorly understood. In this work, we investigate the triangular-lattice Hubbard model with particle doping and provide compelling evidence for an intrinsic, interaction-driven quantum state, which we term the Nagaoka supermetal. This state is characterized by a sublinear temperature dependence in the DC resistivity, along with singular behaviors in the charge compressibility and zero-frequency spectral weight. To understand the origin of these singular properties, we derive an effective low-energy model and demonstrate that a higher-order Van Hove singularity emerges from the reconstructed dispersion. This singularity gives rise to a power-law divergence in the density of states, capturing the anomalous properties observed in the supermetallic regime. Our findings offer a new perspective on non-Fermi liquid states in geometrically frustrated systems and are directly accessible in current ultracold atom experiments.
\end{abstract}

\date{\today}

\maketitle

{\it Introduction.---} 
Landau’s Fermi liquid theory describes interacting fermions through the quasiparticle concept, a picture that can break down under certain conditions. This collapse manifests in anomalous phases such as strange metals~\cite{annurev090921,Arovas2022,doi:10.1126/science.aau7063,doi:10.1073/pnas.2115819119,RevModPhys.75.1085} and the pseudogap phase~\cite{RevModPhys.78.17,doi:10.1126/science.ade9194} in the presence of strong electronic correlations. Another intriguing non-Fermi liquid (NFL) addition to this frontier is the recently proposed supermetal—a state instead originating from a higher-order Van Hove singularity (HOVHS) featuring a power-law divergent density of states (DOS)~\cite{annurev042924, PhysRevResearch.1.033206, yuan2019magic,chandrasekaran2024engineering,PhysRevLett.123.207202,PhysRevLett.131.026601}.
The supermetal hosts multiple competing divergent susceptibilities without long-range order and exhibits sublinear resistivity $\rho \sim T^{\alpha}$ ($\alpha < 1$), marking a fundamental departure from the standard quasiparticle paradigm.
While pioneering experimental studies on solid-state platforms, such as Kagome metals~\cite{kang2022twofold,hu2022rich}, have unveiled the presence of HOVHS, the direct observation of a supermetallic state has yet to be achieved.

Complementing solid-state systems, ultracold fermionic atoms in optical lattices provide a pristine platform for simulating strongly correlated physics, where quantum gas microscopy enables site-resolved access to many-body correlations~\cite{PhysRevLett.114.213002,PhysRevLett.115.263001,PhysRevLett.114.193001,haller2015single,doi:10.1126/science.aad9041,PhysRevLett.116.235301,doi:10.1126/science.aag1430,mazurenko2017cold,doi:10.1126/science.ade4245,xu2025neutral}.
Beyond half filling, this capability has revealed the microscopic structure of doped carriers, including the formation of magnetic polarons~\cite{koepsell2019imaging,doi:10.1126/science.aav3587,doi:10.1126/science.abe7165,PhysRevX.11.021022,prichard2025magnon} and the emergence of pseudogap signatures~\cite{chalopin2024probing,kendrick2025pseudogap}. 
While much attention has focused on square lattices, the triangular geometry introduces geometric frustration that suppresses conventional antiferromagnetic correlations.   
This frustrated background amplifies the role of doping, facilitating itinerant ferromagnetic correlations at a kinetic energy scale $t$ for finite $U$—far exceeding the superexchange energy scale $J \sim t^2/U$.
Indeed, this kinetic-driven regime has enabled the direct observation of Nagaoka ferromagnetic polarons in doped triangular lattices~\cite{xu2023frustration,lebrat2024observation,prichard2024directly,dehollain2020nagaoka,ciorciaro2023kinetic,qiao2025kinetically,martin2026measuring}, sparking extensive theoretical investigations~\cite{PhysRevA.110.L021303,PhysRevB.109.165131,1k49-msfx,PhysRevB.109.235128,morera2024itinerant,chen2025geometric,PhysRevB.110.L100406,dieplinger2024itinerant,PhysRevB.111.245120,currie2025numerical,reinmoser2025optimized,v3fr-gch1,PhysRevB.110.L121109,10.21468/SciPostPhys.17.3.072}.
However, while the magnetic properties are increasingly well-characterized, how this itinerant magnetism interplays with strong correlations to govern low-energy excitations and charge transport remains an open question.

In this Letter, we investigate the doping physics of strongly interacting fermions on the two-dimensional triangular lattice. 
We provide evidence for a correlated metallic state in the particle-doped regime—the intrinsic Nagaoka supermetal (NS)—as revealed by our cluster dynamical mean-field theory calculations, benchmarked against quantum Monte Carlo and experimental observations.
This state exhibits sublinear DC resistivity and singular behaviors in the charge compressibility and zero-frequency spectral weight.
Distinct from supermetals that rely on fine-tuning of the single-particle band structure~\cite{PhysRevResearch.1.033206}, the NS state here emerges intrinsically from an interaction-driven HOVHS. And the divergence exponents of this NS state evolve systematically with the doping level.
To uncover the microscopic origin of this singularity, we derive an effective model demonstrating that strong correlations induce an emergent next-nearest-neighbor (NNN) hopping mediated by Nagaoka polarons.
This reconfiguration drives the system into the HOVHS regime, where the resulting power-law divergence in the DOS underpins the singular thermodynamic behavior and correlates with the transport anomalies.
Importantly, our analysis shows that the key features of the NS state remain robust within experimentally accessible temperature and interaction regimes.
These findings establish a microscopic route to interaction-driven HOVHS via the Nagaoka mechanism, providing a distinct framework for understanding emergent NFL behavior in frustrated quantum systems.

{\it Model and method.---}
We study the $s$-band Fermi-Hubbard model on the triangular lattice, defined by the Hamiltonian
\begin{eqnarray}
\nonumber
H&=&-t \sum _{\langle i,j\rangle,\sigma} (c^{\dagger}_{i\sigma} c_{j\sigma}^{} +{\rm H.c.}) +U \sum_i n_{i\uparrow} n_{i\downarrow}\\
&-&\mu \sum_i \left(n_{i\uparrow}+n_{i\downarrow}\right),
\label{eq1}
\end{eqnarray}
where $c^{\dagger}_{i\sigma}\left({c}_{i\sigma}\right)$ denotes the fermionic creation (annihilation) operator for spin $\sigma$ on site $i$, and $n_{i\sigma}=c^{\dagger}_{i\sigma}c_{i\sigma}$ is the density operator. Here, $t$ denotes the nearest-neighbor (NN) hopping amplitude, $U$ represents the onsite repulsive interaction, and $\mu$ is the chemical potential.  
In the strongly correlated regime, this frustrated system realizes a Mott-insulating state at half filling and evolves upon doping into multiple distinct regimes, including the Nagaoka ferromagnetism~\cite{PhysRevB.73.235107,lebrat2024observation,prichard2024directly,xu2023frustration}, pseudogap (PG)~\cite{PhysRevB.110.L121109,PhysRevB.79.115116,PhysRevResearch.2.033067}, Fermi liquid (FL)~\cite{PhysRevB.88.041103}, and strange metal (SM)~\cite{10.21468/SciPostPhys.17.3.072}. 
Motivated by recent experimental realizations of this Nagaoka state~\cite{lebrat2024observation,prichard2024directly,xu2023frustration}, we focus on the particle-doped regime and examine the transport, thermodynamic, and spectroscopic signatures arising from the interplay between kinetic magnetism and charge motion.  
In the following, $t$ is taken as the unit of energy.

\begin{figure}[h!]
\includegraphics[trim = 0mm 0mm 0mm 0mm, clip=true, width=\columnwidth]{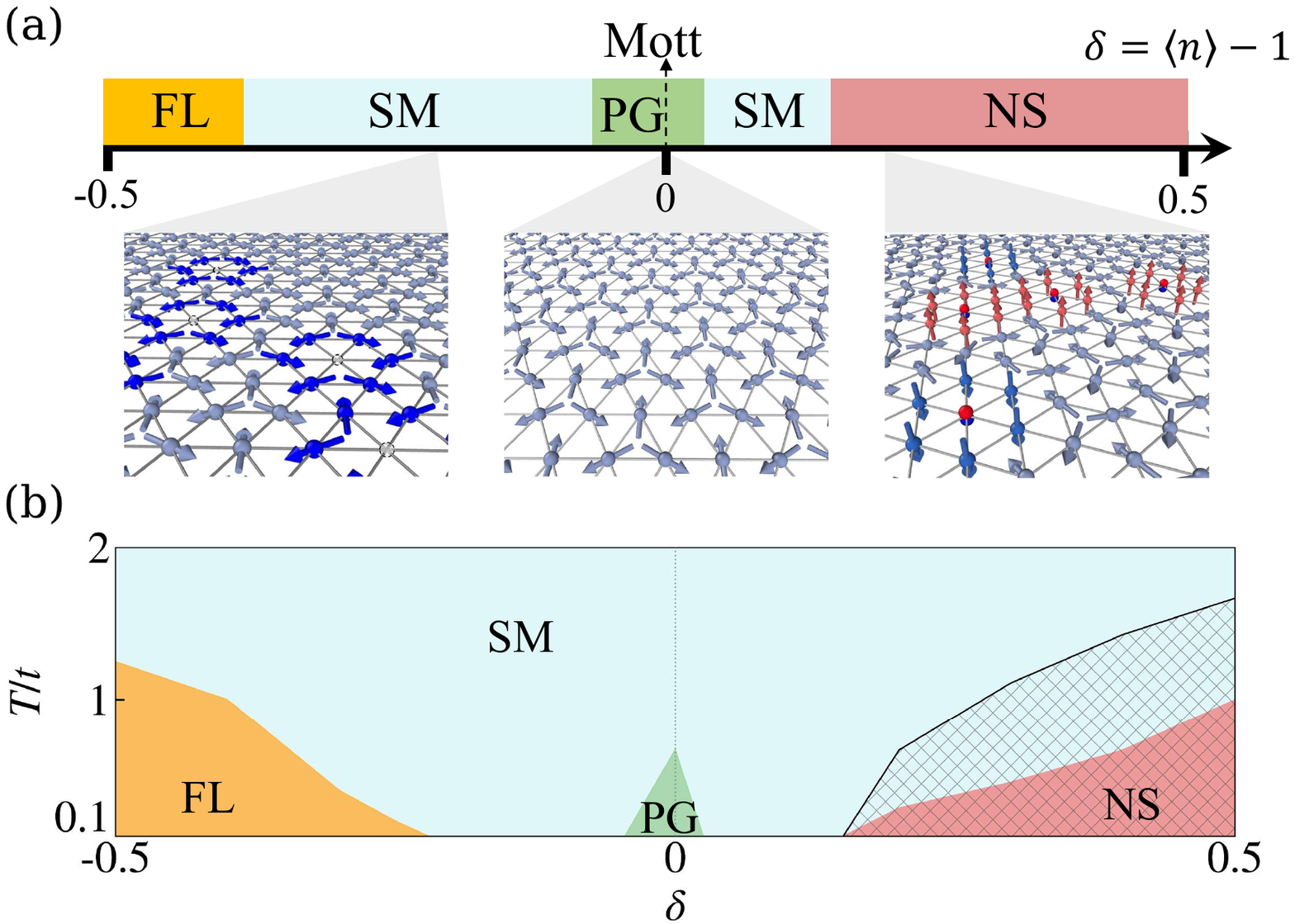}
\caption{Emergence of a Nagaoka supermetal (NS) state in the particle-doped triangular lattice.
(a) Schematic of low-temperature phases mapped by distinct transport, thermodynamic, and spectroscopic signatures. Under hole doping, the system exhibits a crossover from a Mott insulator to a Fermi liquid (FL) through intermediate pseudogap (PG) and strange metal (SM) regimes. In contrast, on the particle-doped side, the SM behavior gives way to a NS state that emerges alongside Nagaoka ferromagnetic correlations at high doping. Bottom panel: Schematic of the corresponding magnetic order.
(b) Finite-temperature phase diagram. The NS state exhibits remarkable robustness, persisting up to $T/t \approx 1$. Cross-hatched regions denote positive nearest-neighbor magnetic correlations, and background colors represent distinct regimes. The interaction strength is set to $U/t=12$.
}
\label{Figure1}
\end{figure}

We solve the model using the dynamical cluster approximation (DCA)~\cite{RevModPhys.77.1027,PhysRevB.61.12739}, which generalizes the dynamical mean-field theory (DMFT) by resolving the momentum dependence of the self-energy. 
By partitioning the Brillouin zone into patches, DCA restores the short-range spatial correlations discarded in the local limit of single-site DMFT, thereby capturing the interplay between onsite dynamical fluctuations and non-local spatial effects. 
The resulting cluster impurity problem is addressed via the continuous-time hybridization expansion~\cite{RevModPhys.83.349} within the TRIQS library~\cite{PARCOLLET2015398}.
We validate our theoretical framework by benchmarking DCA results against Determinant Quantum Monte Carlo (DQMC) simulations~\cite{Blankenbecler1981,Hirsch1983,White1989,Assaad2008} and experimental measurements of local correlations~\cite{lebrat2024observation,prichard2024directly,xu2023frustration}.
These agreements confirm that the relevant many-body physics is captured within our cluster approach (see Supplementary Material for methodology and benchmarks~\cite{SM}).

\begin{figure*}[t]
\includegraphics[trim = 0mm 0mm 0mm 0mm, clip=true, width=1\textwidth]{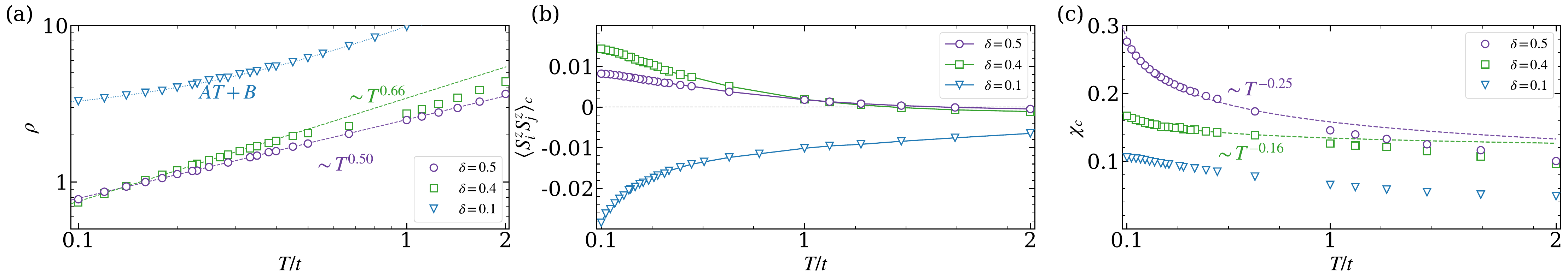} 
\caption{Signatures of the Nagaoka-driven NS state. Temperature dependence of (a) DC resistivity $\rho$, (b) NN spin correlation $\langle S^z_i S^z_j \rangle_c$, and (c) charge compressibility $\chi_c$ for various doping $\delta$. 
The NS state is characterized by sublinear $\rho \sim T^{\alpha}$ with $\alpha < 1$ [dashed lines in (a)] and power-law divergent $\chi_c \sim T^{-\gamma}$ with $\gamma>0$ [dashed lines in (c)]. 
These features emerge as Nagaoka ferromagnetic correlations develop (b), signaling a distinct departure from the high-temperature regime characterized by $\rho = AT+B$ [dotted line in (a)] and antiferromagnetic correlations.
Pronounced transport and thermodynamic anomalies manifest within the regime of well-developed ferromagnetic correlations, consistent with a Nagaoka-driven origin of the NS state. The interaction strength is set to $U/t=12$.}
\label{Figure2}
\end{figure*}

We characterize the magnetic properties using the NN spin correlation function
\begin{eqnarray}
\langle {S}^z_i {S}^z_j \rangle_c=\langle {S}^z_i {S}^z_j \rangle -\langle {S}^z_i \rangle \langle {S}^z_j \rangle,
\end{eqnarray}
where ${S}^z_i=\frac{1}{2} \left({n}_{i\uparrow}-{n}_{i\downarrow}\right)$ is the spin operator.
The system's transport properties are characterized by the DC resistivity $\rho = \lim_{\omega \to 0} [1/\sigma(\omega)]$, computed for various dopings $\delta = \langle n \rangle - 1$.  
Within the DCA framework, the optical conductivity $\sigma(\omega)$ is given by~\cite{PhysRevB.102.115142,PhysRevX.12.021064,RevModPhys.68.13}
\begin{eqnarray}
\nonumber
\sigma (\omega)&=&\sigma_0 \iint d\varepsilon d\nu X(\varepsilon) \mathcal{A}(\varepsilon, \nu) \mathcal{A}(\varepsilon, \nu+\omega)\\
&\times& \frac{n_f(\nu)-n_f(\nu+\omega)}{\omega},
\end{eqnarray}
where $X(\varepsilon) = \frac{1}{N} \sum_{\mathbf{k}} v_{\mathbf{k},x}^2 \delta(\varepsilon - \varepsilon_{\mathbf{k}})$ denotes the transport density of states, with $v_{\mathbf{k},x} = \partial \varepsilon_{\mathbf{k}} / \partial k_x$ being the group velocity along the $x$-direction. Here, $\mathbf{k}$ is the momentum within the Brillouin zone of the $N$-site triangular lattice, and $\varepsilon_{\mathbf{k}}$ represents the bare dispersion. $n_f(\nu)$ represents the Fermi-Dirac distribution, $\sigma_0 = 4\pi / \sqrt{3}$ for a triangular lattice, and $\mathcal{A}(\varepsilon, \nu) = -\frac{1}{\pi} \text{Im} [\nu + \mu - \varepsilon - \Sigma(\nu)]^{-1}$. The real-frequency retarded self-energy $\Sigma(\nu)$ is obtained via the Padé analytical continuation method~\cite{vidberg1977solving}, with results cross-validated using the maximum entropy method~\cite{SM}. 

To characterize the thermodynamic and spectroscopic properties of the system, we evaluate the charge compressibility $\chi_c=\frac{\partial n}{\partial \mu}$ and the zero-frequency spectral weight $A(\omega=0)$~\cite{PhysRevX.11.041013}, respectively. 
$A(\omega=0)$ is extracted by extrapolating the Matsubara Green's function $G(i\omega_n)$ from the lowest frequencies to zero frequency as
\begin{equation}
A(\omega=0)=-\frac{1}{\pi} \lim_{\omega_n \rightarrow 0^+} {\rm Im} G(i\omega_n).
\end{equation}

{\it Evidence for Nagaoka supermetal state.---} 
We reveal a robust Nagaoka supermetal state in the particle-doped triangular lattice [Fig.~\ref{Figure1}(a)]. Driven by intrinsic particle-hole asymmetry, hole doping ($\delta < 0$) is dominated by $120^\circ$ N\'{e}el correlations~\cite{PhysRevLett.95.087202,v3fr-gch1,PhysRevB.110.L041117}, whereas the particle-doped side ($\delta > 0$) supports a Nagaoka ferromagnetic state~\cite{lebrat2024observation,prichard2024directly,morera2024itinerant,xu2023frustration}. 
These distinct magnetic backgrounds result in qualitatively different transport properties.
In the particle-doped regime, the Nagaoka magnetic correlations give rise to the NS state, 
where the resistivity exhibits a sublinear temperature dependence.
This behavior departs from the $T$-linear scaling of strange or bad metals~\cite{doi:10.1126/science.aau7063,doi:10.1126/science.aat4134,PhysRevLett.110.206402,doi:10.1073/pnas.2115819119,eom2025strange,Song2025B,Lu2026}, and is reminiscent of recently proposed supermetallic states~\cite{annurev042924, PhysRevResearch.1.033206, PhysRevLett.131.026601}. 
It suggests an interaction-induced singularity, pointing to the mechanism discussed below.

The thermal stability of the NS state is captured in the finite-temperature phase diagram [Fig.~\ref{Figure1}(b)].  
Unlike conventional correlated phases controlled by the superexchange energy scale $J \sim t^2/U$, the NS regime remains robust up to temperatures set by the kinetic scale $t$.
This energy hierarchy originates from the Nagaoka mechanism, where the underlying ferromagnetic correlation is governed by $t$ rather than $J$.
The elevated scale places the phenomenon within the reach of current quantum simulation experiments~\cite{xu2025neutral,kendrick2025pseudogap,prichard2024directly,chalopin2024probing,mazurenko2017cold,xu2023frustration}.
In addition to the NS state, the system exhibits PG, FL, and SM regimes, each characterized by distinct transport and spectroscopic signatures~\cite{SM}.

To quantitatively characterize the NS regime, we examine the temperature-dependent resistivity $\rho$ across doping $\delta$, highlighting its link to the underlying magnetic correlations.
At high doping ($\delta = 0.5$ and $0.4$), the system enters the NS regime at low temperatures, where $\rho$ exhibits a power-law behavior as $\rho \sim T^{\alpha}$ with $\alpha\approx0.5$ ($\delta=0.5$) and $\alpha\approx0.66$ ($\delta=0.4$) [Fig.~\ref{Figure2}(a)].
As the temperature increases, the system undergoes a crossover and $\rho$ recovers its standard linear-in-$T$ dependence at high temperatures~\cite{Song2025B,Lu2026}.
Crucially, this crossover is closely tied to the magnetic environment: the transport anomaly emerges alongside the buildup of positive NN spin correlations [Fig.~\ref{Figure2}(b)], indicating that ferromagnetic correlations drive the onset of sublinear scaling.
This interpretation is further corroborated by the low-doping case ($\delta = 0.1$): here, NN spin correlations remain antiferromagnetic and no crossover occurs, with $\rho$ retaining linearity across the entire temperature range.
The extracted exponent is robust against finite-size effects, as confirmed by cluster-size analysis~\cite{SM}.

Further evidence for the NS state is provided by the charge compressibility $\chi_c$ [Fig.~\ref{Figure2}(c)]. At low temperatures, $\chi_c$ exhibits a power-law divergence, $\chi_c \propto T^{-\gamma}$ with $\gamma \approx 0.25$ ($\delta=0.5$) and $\gamma \approx 0.16$ ($\delta=0.4$), reflecting the singular DOS associated with the NS regime.
Crucially, the supermetal theory~\cite{PhysRevResearch.1.033206} predicts a fundamental scaling relation between the compressibility and resistivity exponents, namely $\chi_c \propto T^{-\gamma}$ and $\rho \propto T^{\alpha}$ with $\alpha = 1-2\gamma$. Our numerical results at doping $\delta=0.5$ show excellent agreement with this prediction, where the scaling exponents $\alpha \approx 0.5$ and $\gamma \approx 0.25$ satisfy the relation precisely. Similarly, at $\delta = 0.4$, the exponents $\alpha \approx 0.66$ and $\gamma \approx 0.16$ also approximately obey this relation. This quantitative agreement between thermodynamic [Fig.~\ref{Figure2}(c)] and transport signatures [Fig.~\ref{Figure2}(a)] provides a self-consistent verification of the NS state.
At $\delta = 0.1$, although $\chi_c$ also increases with decreasing temperature, it deviates from the aforementioned power-law divergence.
Notably, our numerical results are consistent with DQMC simulations at elevated temperatures where the fermion sign problem remains tractable~\cite{SM}, even without long-wavelength vertex corrections in DCA (see discussions in Sec. III in the Supplementary Material).

\begin{figure}[t!]
\includegraphics[trim = 0mm 0mm 0mm 0mm, clip=true, width=\columnwidth]{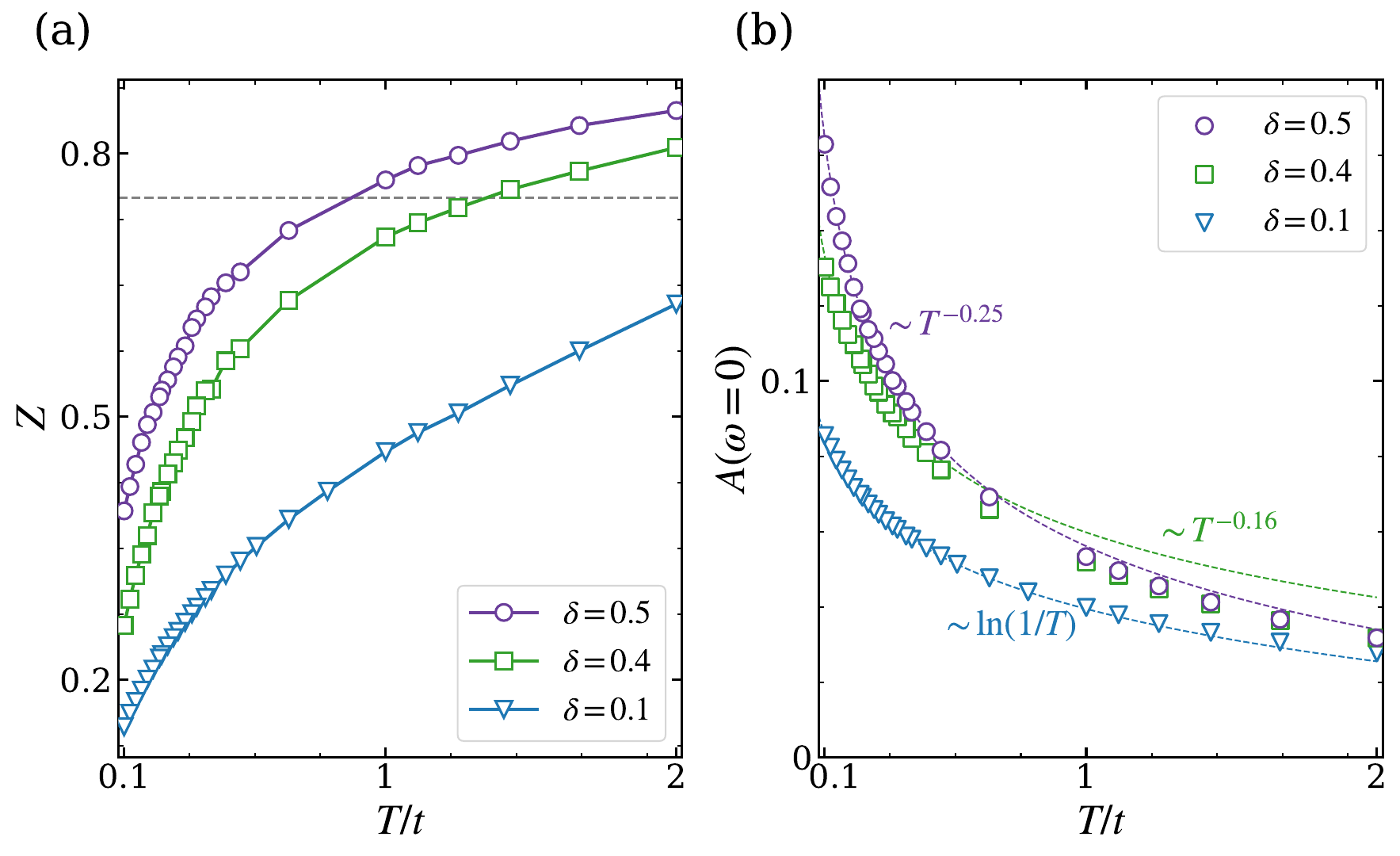}
\caption{Characteristics of the Nagaoka-driven HOVHS in the particle-doped regime. (a) Kinetic renormalization factor $Z$ as a function of doping and temperature.
The dashed line denotes the $Z=0.75$ contour, marking the onset of the interaction-induced HOVHS and coinciding with the emergence of sublinear resistivity [Fig.~\ref{Figure2}(a)].
(b) Zero-frequency spectral weight $A(\omega=0)$ as a function of doping and temperature.
The dashed lines represent the fits to their corresponding low-temperature data sets. The interaction strength is set to $U/t=12$.}
\label{Figure3}
\end{figure}

{\it Effective model.---}
To elucidate the microscopic origin of the NS state, we analytically map the Hubbard model onto a low-energy effective theory using second-order perturbation theory. In the particle-doped regime, the underlying physics is dominated by doublon dynamics within a polarizable spin background. In particular, the second-order virtual excitations give rise to correlated three-site hopping processes, which are essential for capturing the singular properties of the triangular lattice. The resulting effective Hamiltonian, projected onto the low-energy subspace, is given by~\cite{SM}
\begin{eqnarray}
H_{\text{eff}} &\sim&  \sum_{\langle i,j \rangle}\tilde{t} d^{\dagger}_{i} d_{j} +  \sum_{\langle\langle i,k \rangle\rangle} t' d^{\dagger}_{i}d_{k} + {\rm H.c.},
\label{effective}
\end{eqnarray}
where $d^{\dagger}_i$ denotes the doublon creation operator, with the singly occupied spin background acting as the effective vacuum.
$\tilde{t}=Zt$ represents the renormalized NN hopping amplitude, with the kinetic renormalization factor 
$Z$ capturing the effect of many-body correlations. $t' = \frac{2t^2}{U}(\frac{1}{2}-m)$ is the effective NNN hopping, arising from the correlated three-site processes, where $m=\left|\left\langle S^z_j \right\rangle\right|$ denotes the magnitude of the background magnetization. 

While the $\mathbf{M}$ point in the Brillouin zone traditionally hosts the standard Van Hove singularity of the triangular lattice for $\delta=0.5$, it here serves as the starting point for the interaction-induced band reconfiguration. By expanding the dispersion relation of Eq.~(\ref{effective}) around the $\mathbf{M}$ point, we obtain the low-energy dispersion $(\left|{\mathbf{q}}\right| \to 0)$
\begin{eqnarray}
\varepsilon(\mathbf{M}+\mathbf{q})=-\frac{1}{2}(\tilde{t}-9t')q^2_x+\frac{3}{2}(\tilde{t}-t')q^2_y+ {\mathcal O}(q^4).
\label{dispersion}
\end{eqnarray}
Eq.~(\ref{dispersion}) shows that, in the vicinity of $U/t \approx 10$, the leading quadratic term along $q_x$ is strongly suppressed by interaction-driven renormalization, resulting in an approximate cancellation of the $q_x^2$ coefficient ($\tilde{t} \approx 9t'$). This band reconstruction yields a low-energy dispersion approaching the quartic form $\varepsilon(\mathbf{M}+\mathbf{q}) \sim q_x^4 - q_y^2$, characteristic of a HOVHS.
Such a reconstruction is reminiscent of the interaction-driven band flattening near the Van Hove point at $\mathbf{M}$ reported in Ref.~\cite{PhysRevLett.112.070403}, supporting a consistent microscopic scenario for the emergence of a HOVHS. 
Quantitatively, realizing this interaction-induced HOVHS at $\delta=0.5$ requires the renormalization factor $Z$ to satisfy the condition $0 < Z < 9t/U$. For a representative $U/t=12$, this condition restricts $Z$ to the range $(0, 3/4)$.
The evolution of $Z$ as a function of doping and temperature is presented in Fig.~\ref{Figure3}(a). Notably, the onset temperature where $Z$ enters the HOVHS regime matches the transport crossover temperature observed in the resistivity [Fig.~\ref{Figure2}(a)].

The existence of the HOVHS is further supported by the DOS extracted via the single-particle spectral weight $A(\omega=0)$~\cite{Coleman2015}. As shown in Fig.~\ref{Figure3}(b), $A(\omega=0)$ in the NS regime exhibits a robust power-law divergence scaling as $T^{-0.25}$ for $\delta=0.5$ and $T^{-0.16}$ for $\delta=0.4$.
Notably, this low-temperature divergence follows a $T^{-0.25}$ scaling at $\delta=0.5$, precisely matching the  $|\delta E| ^{-1/4}$ singularity expected for a high-order Van Hove point in the noninteracting limit (with energy $\delta E$ relative to the singularity)~\cite{SM}.
Moreover, the excellent consistency between the scaling exponents of the spectral weight and charge compressibility [Fig.~\ref{Figure2}(c)] provides strong evidence identifying the interaction-induced HOVHS as the microscopic origin of the NS state.
In contrast, at $\delta = 0.1$, $A(\omega=0)$ exhibits a $\ln(1/T)$ scaling, reminiscent of the marginal Fermi liquid behavior~\cite{PhysRevB.45.5714,PhysRevB.46.11798,PhysRevResearch.1.033206}. Furthermore, the absence of the power-law signature elsewhere in the hole-doped regime~\cite{SM} highlights the specificity of the HOVHS-driven features in the NS state. 

\begin{figure}[t!]
\includegraphics[trim = 0mm 0mm 0mm 0mm, clip=true, width=\columnwidth]{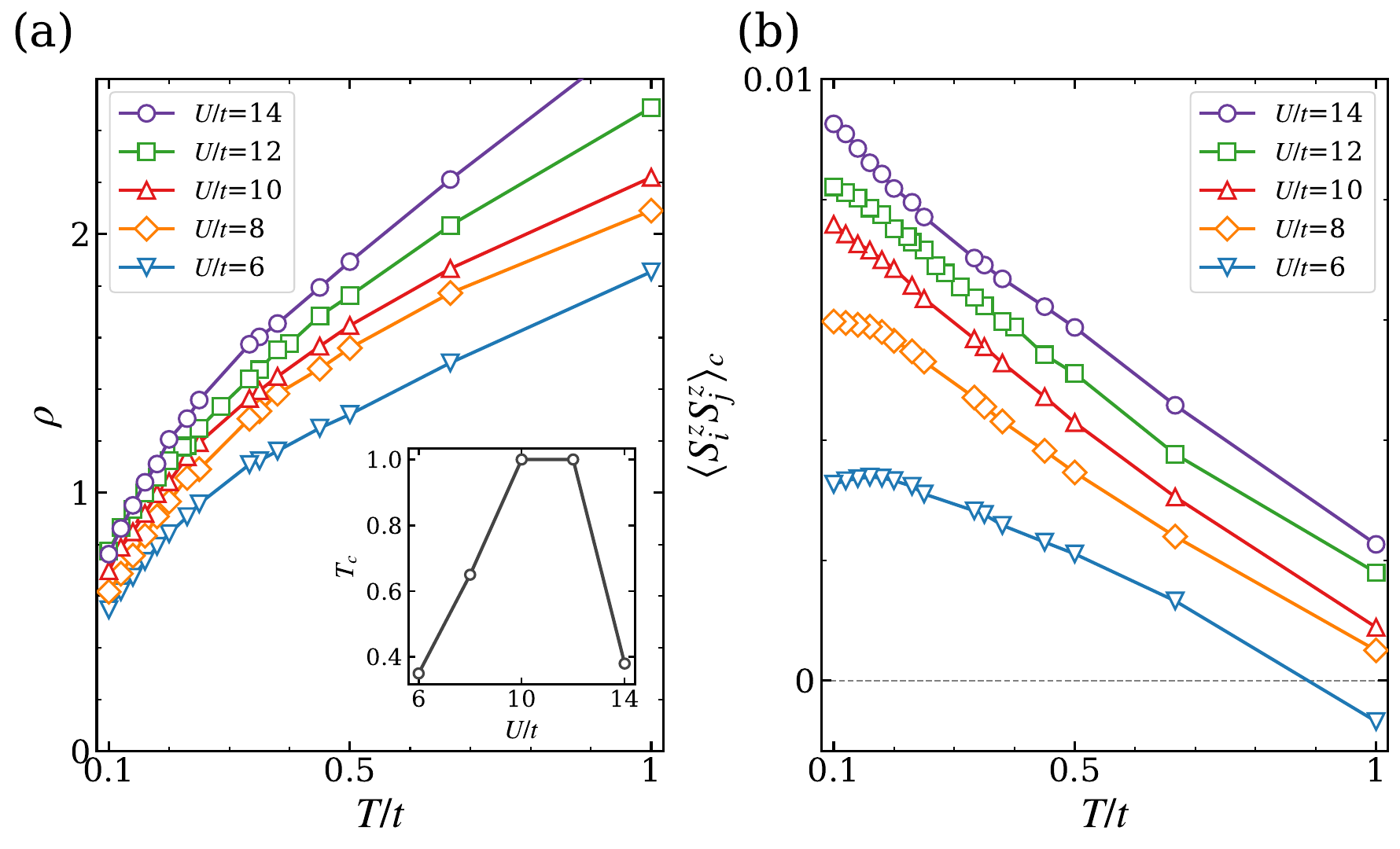}
\caption{Impact of interaction strength $U$ on transport and magnetic properties. (a) DC resistivity $\rho$ as a function of temperature $T/t$ for different interaction strengths $U/t$. The inset shows the onset temperature $T_c$ of the crossover into the NS regime, extracted from $\rho$. (b) Temperature dependence of the NN spin correlation function. The sign-change temperature shifts toward lower $T$ as $U/t$ decreases. The doping concentration is fixed at $\delta=0.5$.}
\label{Figure4}
\end{figure}

{\it Interaction dependence of Nagaoka supermetal.---}
Finally, we investigate the influence of interaction strength $U$ on the stability of the NS state. As shown in Fig.~\ref{Figure4}(a), the sublinear transport remains robust across a wide range of $U$, maintaining a behavior entirely distinct from SM and FL regimes. Notably, the characteristic onset temperature $T_c$, deduced from the resistivity crossover, displays a prominent non-monotonic dependence on $U$ [Fig.~\ref{Figure4}(a), inset], with a maximum at intermediate coupling. 

This non-monotonicity stems from the competition between two distinct physical limits. At large $U$, strong correlations suppress the effective NNN hopping, reducing the temperature window for the NS state.
Conversely, at small $U$, the formation scale of Nagaoka ferromagnetic polarons is lowered, making the NS state more susceptible to thermal fluctuations [Fig.~\ref{Figure4}(b)].
Furthermore, at elevated temperatures, a linear-in-$T$ resistivity becomes increasingly prominent, marking a crossover into the SM regime across all $U$.

{\it Conclusion and outlook.---}
In summary, we have investigated the transport properties of the triangular-lattice Hubbard model in the particle-doped regime and identified the emergence of a Nagaoka supermetal state at low temperatures. This state is characterized by power-law divergences in the charge compressibility and zero-frequency spectral weight, and a sublinear temperature dependence in the DC resistivity. 
Distinct from conventional supermetals rooted in the single-particle picture~\cite{PhysRevResearch.1.033206}, this state is intrinsic and many-body in nature. It emerges from an interaction-induced higher-order Van Hove singularity and features doping-dependent divergence exponents.
These theoretical predictions are directly accessible to current quantum simulation experiments via measurements of diffusion and compressibility~\cite{doi:10.1126/science.aat4134,kendrick2025pseudogap,lebrat2025ferrimagneti}.
Looking forward, extending this framework to spin-polarized regimes~\cite{prichard2025magnon,qiao2025kinetically} will clarify how collective spin excitations dress charge carriers and reshape the transport landscape. Additionally, investigating transport signatures within the quantum spin liquid phase~\cite{RevModPhys.89.025003} stands as a compelling direction for future research.

We acknowledge useful discussions with Hui Zhai, Chengshu Li, Wei Wu, and Rongqiang He. This work was supported by the National Natural Science Foundation of China (under Grants No. 12374252, No. 12074431, No. 12174130, No. 12304076, No. 12247103, and No. 12204377), the Excellent Youth Foundation of Hunan Scientific Committee under Grant No. 2021JJ10044, the Quantum Science and Technology-National Science and Technology Major Project (Grant No. 2021ZD0301900), and the Youth Innovation Team of Shaanxi Universities. Numerical calculations were performed on the TianHe-1A cluster of the National Supercomputer Center at Tianjin.

\let\oldaddcontentsline\addcontentsline
\renewcommand{\addcontentsline}[3]{} 
\bibliography{references.bib}
\let\addcontentsline\oldaddcontentsline

\clearpage       
\onecolumngrid

\begin{center}
\textbf{\large Supplementary Material}
\end{center}
\vspace{2ex}

\setcounter{MaxMatrixCols}{14}
\setcounter{equation}{0}
\setcounter{figure}{0}
\setcounter{table}{0}

\makeatletter
\renewcommand{\theequation}{S\arabic{equation}}
\renewcommand{\thefigure}{S\arabic{figure}}
\renewcommand{\thetable}{S\arabic{table}}
\makeatother

\tableofcontents
\vspace{2em}

\section{Geometric Configurations of the DCA Clusters}
The results presented in this study are primarily obtained using the dynamical cluster approximation (DCA)~\cite{RevModPhys.77.1027,PhysRevB.61.12739}, a cluster extension of dynamical mean-field theory (DMFT). In the DCA formalism, the infinite lattice is mapped onto a finite cluster of size $N_c$, which effectively coarse-grains the first Brillouin zone into $N_c$ patches. Each patch is centered around a cluster momentum $\mathbf{Q}$, which is conjugate to the intra-cluster spatial coordinates, thereby leading to a momentum-dependent self-energy. This explicit momentum dependence is a crucial advancement over single-site DMFT, whose local self-energy inherently restricts it to capturing only local quantum fluctuations. Consequently, in addition to local quantum fluctuations, the DCA formalism explicitly captures short-range spatial correlations within the cluster, while treating longer-range correlations at the mean-field level. This capability renders it particularly advantageous for investigating doped systems and calculating multi-site correlation functions. 

\begin{figure}[h!]
\includegraphics[width=0.8\textwidth]{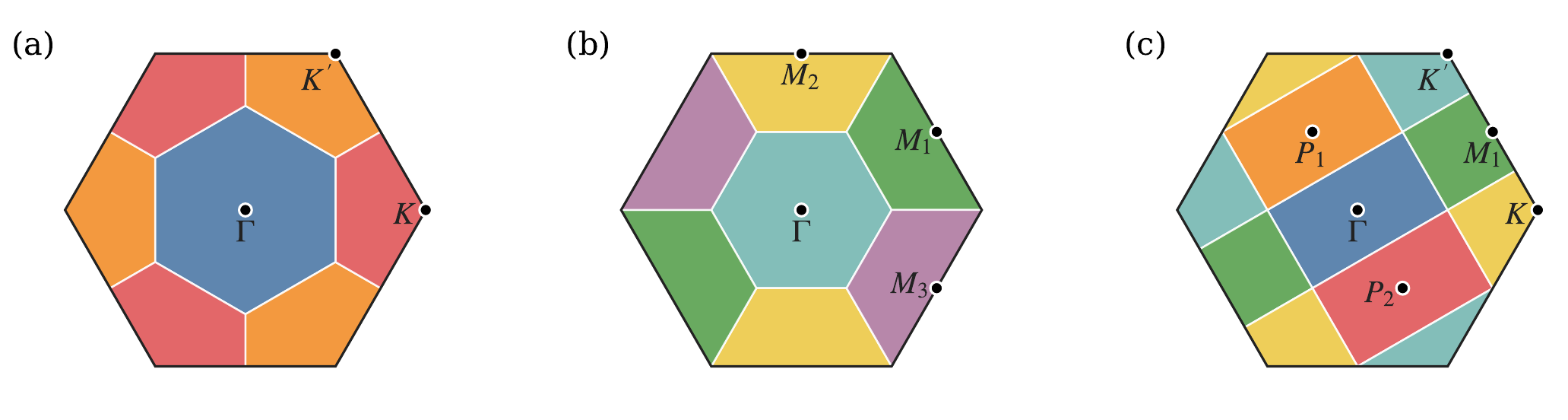}
\caption{Momentum-space coarse-graining patches for the (a) 3-site, (b) 4-site, and (c) 6-site clusters used in the DCA simulations.}
     \label{AFig1}
\end{figure}

The effective cluster impurity problem is solved by the continuous-time hybridization expansion solver (CTHYB)~\cite{RevModPhys.83.349}. The geometric configurations of the DCA clusters used in this work are illustrated in Fig.~\ref{AFig1}. Due to the exponential scaling of the CTHYB solver with the cluster size and the severe fermion sign problem on the frustrated triangular lattice, our calculations are restricted to $N_c = 3$, $4$, and $6$. 
The selection of cluster geometry determines the accessible momentum resolution and the specific high-symmetry points included within the coarse-grained Brillouin zone. In particular, the discretized cluster momenta for the 3- and 6-site geometries include the high-symmetry point $K= (\frac{2\pi}{3}, \frac{2\pi}{\sqrt{3}})$, whereas the 4- and 6-site clusters encompass the point $M= (\pi, \frac{\pi}{\sqrt{3}})$. 
Therefore, these momentum patches capture the low-energy fluctuations associated with the relevant wave vectors within the accessible cluster sizes.

\begin{figure}[h!]
\includegraphics[width=0.6\textwidth]{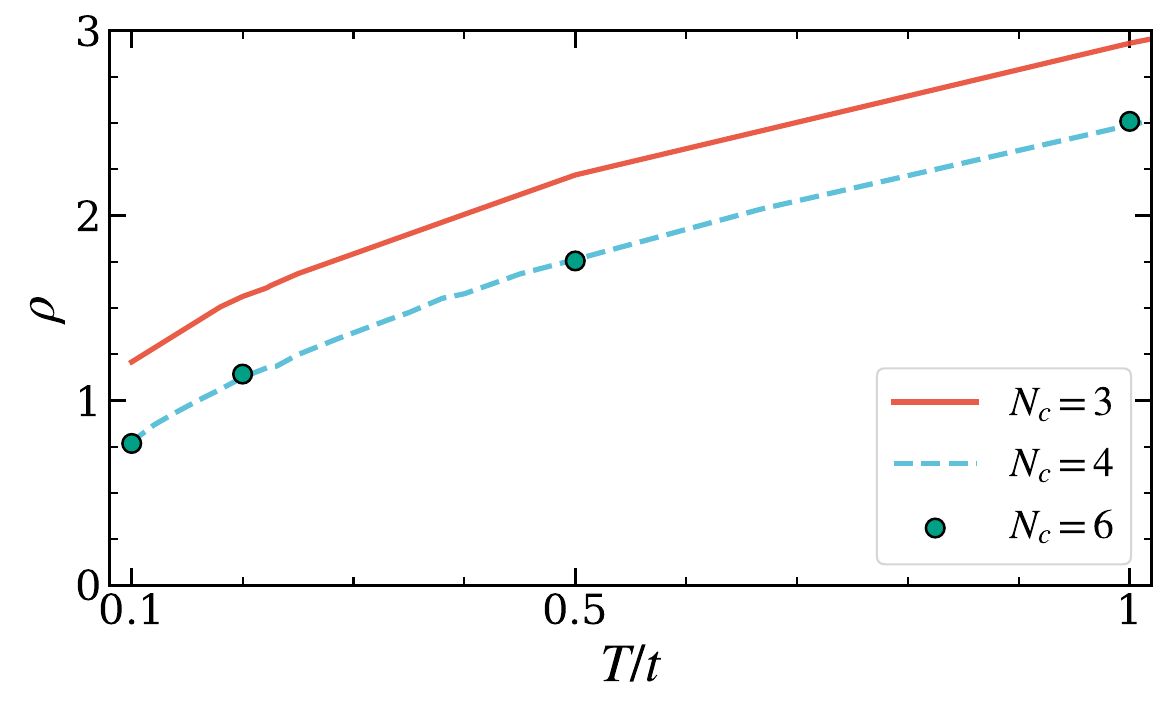}
\caption{Temperature dependence of the DC resistivity for different DCA cluster geometries. $N_c$ denotes the number of coarse-graining patches used in the DCA calculation. Model parameters are $U/t=12$ and $\delta=0.5$.}
     \label{AFig2}
\end{figure}

Furthermore, we examine the dependence of our results on cluster geometry and size, and find that a 4-site cluster—the configuration adopted in the main text—is sufficient to capture the essential physics. As shown in Fig.~\ref{AFig2}, we compare the temperature-dependent resistivity for different cluster configurations in the overdoped regime $\delta=0.5$. In this regime, transport is dominated by the states near the $\mathbf{M}$ point. The 3-site cluster, which lacks momentum resolution at the $\mathbf{M}$ point, yields a higher resistivity. In contrast, both 4-site and 6-site clusters resolve the $\mathbf{M}$ point and exhibit nearly identical resistivity profiles.

\section{Benchmark with experimental data and DQMC simulations}
\begin{figure}[htb!]
\includegraphics[width=1\textwidth]{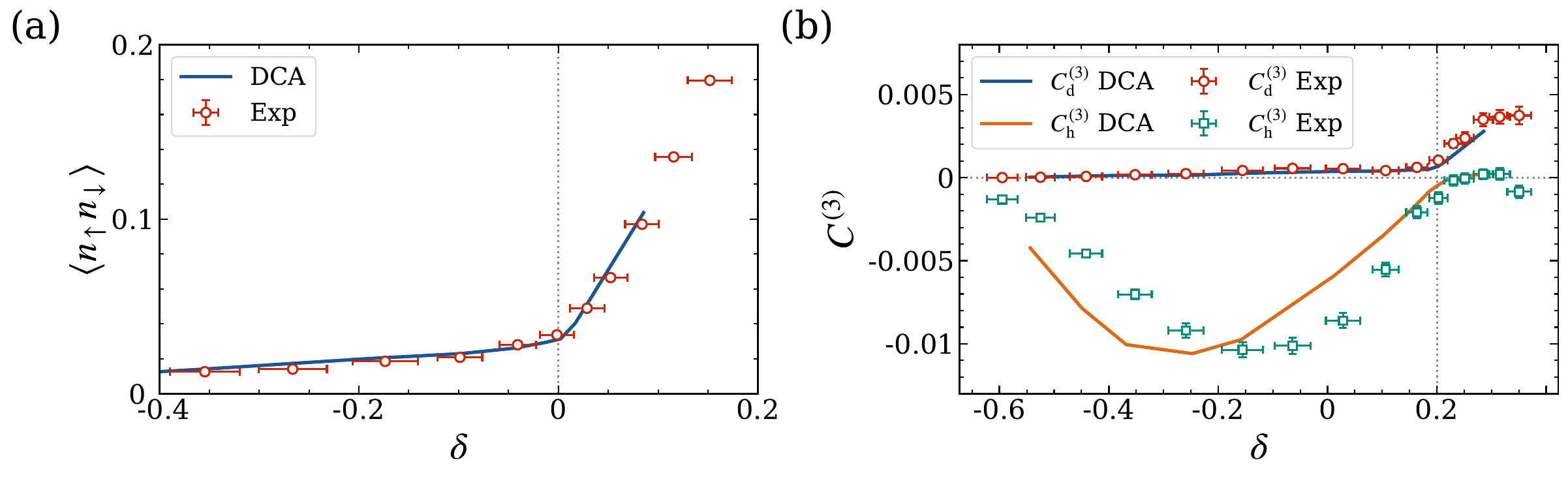}
\caption{Comparison with experimental data. (a) Double occupancy and (b) connected three-site correlations as a function of doping. Dots with error bars denote experimental measurements from Ref.~\cite{prichard2024directly}, while solid lines represent DCA results. Parameters are $U/t=12$ and $T/t=1$.}
     \label{AFig3}
\end{figure}
\begin{figure}[htb!]
\includegraphics[width=1\textwidth]{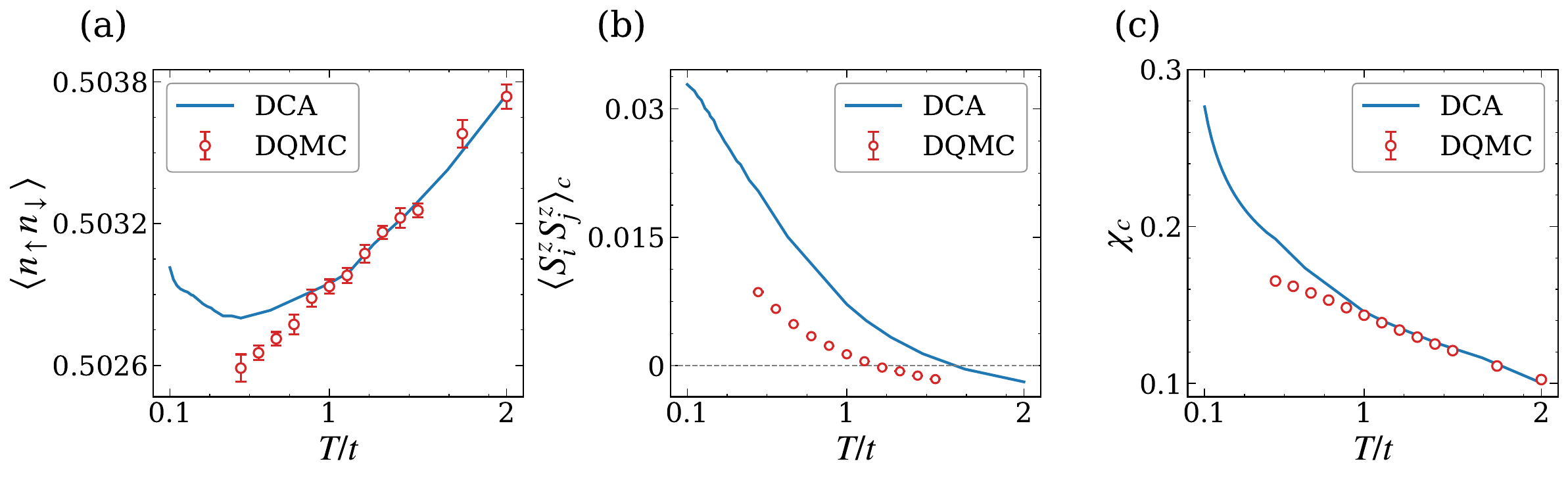}
\caption{Validation against DQMC data. (a) Double occupancy, (b) nearest-neighbor spin correlation function, and (c) charge compressibility as functions of temperature. Dots represent DQMC data from a $9\times9$ system, while solid lines denote DCA calculations. Model parameters are $U/t=12$ and $\delta=0.5$.}
     \label{AFig4}
\end{figure}
To establish the reliability of our DCA framework, we benchmark our calculation results against both experimental measurements and {\it numerically exact} Determinant Quantum Monte Carlo (DQMC) simulations~\cite{Blankenbecler1981,Hirsch1983,White1989,Assaad2008}. We begin by analyzing the double occupancy and three-site correlations, directly comparing our theoretical predictions with experimental measurements~\cite{prichard2024directly}. Specifically, the three-site doublon (hole)-spin-spin correlation function is defined as
\begin{eqnarray}
C^{(3)}_{\rm h/d}\equiv 4\left\langle \left({n}^{\rm h/d}_{i}-\langle {n}^{\rm h/d}_{i} \rangle\right)\left({S}^z_{j} -\langle {S}^z_{j}\rangle \right)\left( {S}^z_{k}-\langle {S}^z_{k}\rangle \right)\right\rangle ,
\end{eqnarray}
where sites $i, j , k$ form a triangle. 
The hole and doublon correlation functions, $C^{(3)}_{\rm h}$ and $C^{(3)}_{\rm d}$, are constructed from the occupancy operators $n^{\rm h}_i = (1-n_{i\uparrow})(1-n_{i\downarrow})$ and $n^{\rm d}_i = n_{i\uparrow}n_{i\downarrow}$, respectively. 
As shown in Fig.~\ref{AFig3}, the DCA framework yields not only quantitative agreement in the double occupancy but also captures the qualitative features of the connected three-site correlations.

Furthermore, to verify the reliability of the DCA in the overdoped regime $\delta=0.5$, we benchmark our results against DQMC simulations performed on an $L=9\times9$ lattice.
As shown in Fig.~\ref{AFig4}, the double occupancy and charge compressibility exhibit quantitative agreement at high temperatures, although slight deviations emerge as the temperature decreases. Regarding the magnetic properties, the DCA accurately captures the qualitative trends of the spin correlations $\langle S^z_i S^z_j \rangle_c$ present in the DQMC data. 
Overall, this consistent agreement across local and non-local observables validates the DCA as a reliable framework for capturing the essential many-body physics in both charge and magnetic degrees of freedom.

\section{Comparison between MaxEnt and Padé Methods}
\begin{figure}[th!]
\includegraphics[width=0.5\textwidth]{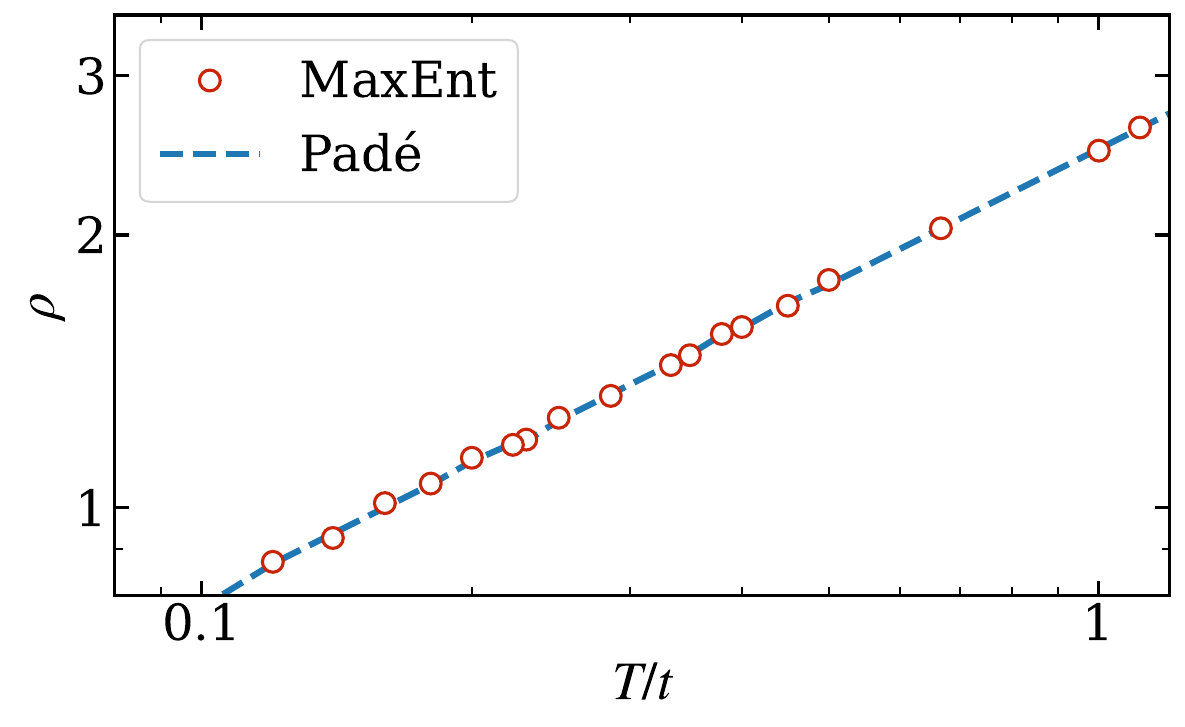}
\caption{Comparison of analytic continuation methods for the DC resistivity $\rho$ at $U/t=12$ and $\delta=0.5$. Results obtained from the MaxEnt and Padé methods are shown as dots and dashed lines, respectively.}
     \label{AFig5}
\end{figure}

The CTHYB solver yields the self-energy $\Sigma(i\omega_n)$ on the Matsubara axis, necessitating analytic continuation to extract the real-frequency self-energy $\Sigma(\nu)$. This procedure is particularly challenging because analytic continuation is an ill-conditioned problem, where statistical noise inherent in the CTHYB data can be severely amplified, leading to unphysical artifacts in the real-frequency spectra. To ensure the reliability of our analytic continuation, we cross-check the results obtained from the Padé approximant~\cite{vidberg1977solving} against those derived using the Maximum Entropy (MaxEnt) method~\cite{PhysRevB.96.155128,JARRELL1996133}. 
As illustrated in Fig.~\ref{AFig5}, the resistivity obtained from both analytic continuation methods exhibits excellent agreement. Owing to its superior low-temperature resolution~\cite{PhysRevB.93.075104,PhysRevX.12.021064}, the Padé method is employed throughout this study. Although vertex corrections are neglected in the resistivity calculation, previous studies have established that the higher connectivity and magnetic frustration of the triangular lattice promote the locality of the self-energy and thereby suppress vertex corrections compared to the square lattice~\cite{PhysRevB.102.115142,PhysRevResearch.2.033434,PhysRevLett.123.036601}. Within this framework, the bubble approximation is thus expected to provide a reliable description of the transport properties, capturing the dominant low-energy behavior even without the explicit inclusion of vertex corrections.

\section{Finite-temperature properties of the Hole-Doped Triangular Lattice}

\begin{figure}[htb!]
\includegraphics[width=1\textwidth]{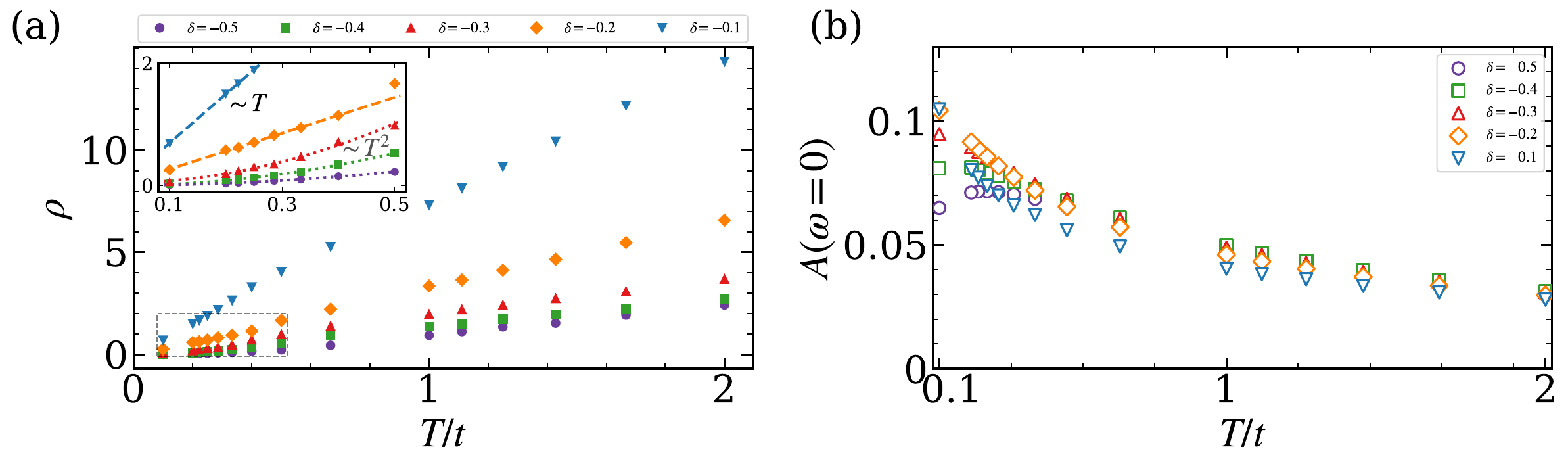}
\caption{
Temperature dependence of the DC resistivity $\rho$ and spectral weight $A(\omega=0)$ for various hole concentrations at $U/t=12$. (a) $\rho$ as a function of $T$
. Inset: Low-temperature scaling, with dashed and dotted lines indicating $\rho \propto T$ and $\rho \propto T^2$, respectively. The legend is shown at the top of the figure. (b) $A(\omega=0)$ versus $T$. The spectral weight exhibits no evidence of power-law divergence for all hole concentrations investigated here.
}
\label{AFig6}
\end{figure}

While the main text primarily investigates the particle-doped triangular lattice to elucidate the Nagaoka supermetal state, here we examine the low-energy physics in the hole-doped regime.
Owing to the lack of particle-hole symmetry in the triangular lattice, hole doping leads to transport behaviors fundamentally different from the particle-doped case. We quantify this distinction by evaluating the temperature-dependent resistivity $\rho$ across various hole concentrations, as summarized in Fig.~\ref{AFig6} (a).
With increasing hole doping, the system evolves from a Mott insulator into a strange metal regime characterized by robust $T$-linear resistivity. Upon further doping, this strange metal eventually crosses over into a conventional Fermi liquid, where the $\rho \propto T^2$ scaling is recovered at low temperatures.
This sharp contrast between the two doping regimes underscores the pivotal role of Nagaoka ferromagnetic polarons in driving the anomalous transport unique to the particle-doped system.

Furthermore, we examine the temperature dependence of the zero-frequency spectral weight $A(\omega=0)$, as shown in Fig.~\ref{AFig6} (b). Notably, within the hole-doped regime, $A(\omega=0)$ exhibits no evidence of power-law divergence for all hole concentrations investigated here. The distinct behavior of the spectral weight indicates the absence of the supermetal state in the hole-doped regime.

\section{Pseudogap phase in the doped triangular lattice}

\begin{figure}[th!]
\includegraphics[width=0.6\textwidth]{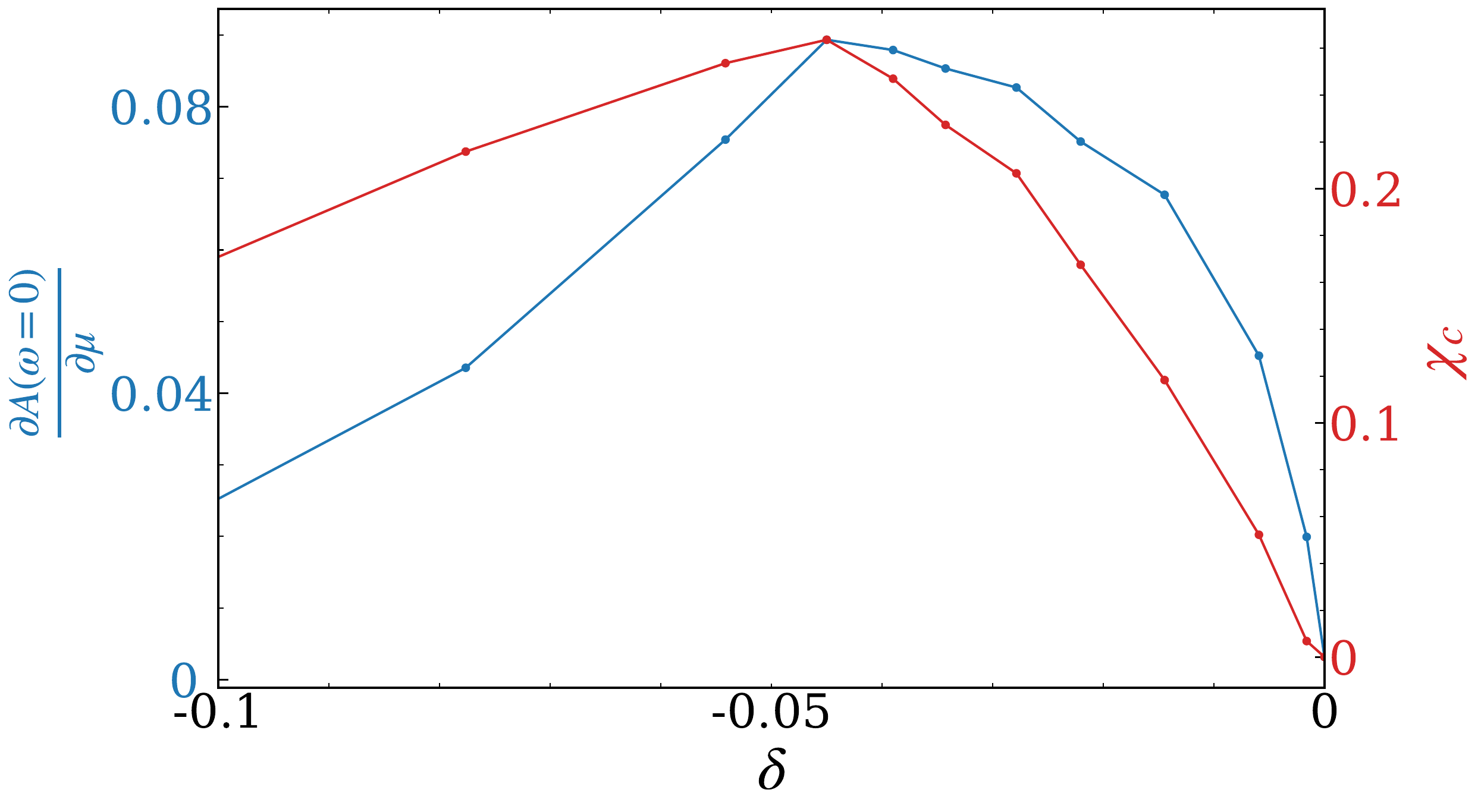}
\caption{Doping dependence of the derivative of the zero-frequency spectral weight $\partial A(\omega=0)/\partial \mu$ and the charge compressibility $\chi_c$ at $U/t=12$. The pseudogap crossover is identified by the maxima in these observables. Data are presented at a fixed temperature $T/t=0.1$.}

\label{AFig8}
\end{figure}

As illustrated in Fig.~1(b) of the main text, the system enters a pseudogap regime at low doping. In this section, we further characterize the crossover into this regime by analyzing both spectroscopic and thermodynamic observables: the local spectral weight $A(\omega=0)$ and the charge compressibility $\chi_c$.
As depicted in Fig.~\ref{AFig8} for $T=0.1t$, the system undergoes a non-monotonic evolution as a function of doping. The crossover from the pseudogap to the strange metal regime is characterized by peaks in both $\chi_c$ and $\partial A(\omega=0)/\partial \mu$. These signatures serve as thermodynamic and spectroscopic markers for the pseudogap boundary, following the criteria established in previous studies~\cite{sordi2012pseudogap,PhysRevLett.108.216401,PhysRevB.93.245147,kendrick2025pseudogap}.

\section{Effective low-energy model for the particle-doped triangular lattice}
To derive the effective low-energy model for the particle-doped case, we divide the total Hilbert space into two orthogonal subspaces. We define the projection operator $P$ and its complement $Q$, where $P$ projects onto the low-energy Hilbert space. The subspace associated with $P$ is spanned by the following states
\begin{eqnarray}
   \mathcal{H}_P=\left\{\left|d;\uparrow;\uparrow\right\rangle, \left|d;\uparrow;\downarrow\right\rangle, \left|d;\downarrow;\uparrow\right\rangle, \left|d;\downarrow;\downarrow\right\rangle, \left|\uparrow;d;\uparrow\right\rangle, \left|\uparrow;d;\downarrow\right\rangle, \left|\downarrow;d;\uparrow\right\rangle, \left|\downarrow;d;\downarrow\right\rangle, \left|\uparrow;\uparrow;d\right\rangle, \left|\uparrow;\downarrow;d\right\rangle, \left|\downarrow;\uparrow;d\right\rangle, \left|\downarrow;\downarrow;d\right\rangle\right\},
\end{eqnarray}
where $\left|d\right\rangle=\left|\uparrow, \downarrow\right\rangle$ denotes a state with double occupancy. The complementary subspace $\mathcal{H}_Q$ associated with $Q$ is spanned by
\begin{eqnarray}
   \mathcal{H}_Q=\left\{ \left|d;0;d\right\rangle, \left|d;d;0\right\rangle, \left|0;d;d\right\rangle \right\}. 
\end{eqnarray}
By applying the projection operators $P$ and $Q$, the eigenvalue equation $H|\Psi\rangle=E|\Psi\rangle$ can be rewritten as
\begin{eqnarray}
\nonumber
\left(QH_tP+QH_tQ+QH_UP+QH_UQ\right)|\Psi\rangle&=&EQ|\Psi\rangle,\\
\left(PH_tP+PH_tQ+PH_UP+PH_UQ\right)|\Psi\rangle&=&EP|\Psi\rangle,
\label{pertubation}
\end{eqnarray}
where $H_U$ and $H_t$ are the interaction and hopping parts of the Hamiltonian, respectively. Solving the first part of Eq.~(\ref{pertubation}) for $Q|\Psi\rangle$ and substituting the result into the second one leads to the second-order effective Hamiltonian
\begin{eqnarray}
    H_{\text{eff}}&=&PH_tP+PH_tQ \frac{1}{E-QH_UQ} QH_tP.
\end{eqnarray}

For the particle-doped case on a triangular lattice, the effective Hamiltonian can be divided into two cases depending on the connectivity between the initial site $i$ and the final site $k$. When $i$ and $k$ are nearest neighbors, the matrix representation of $H_{\text{eff}}$ is given by 
\begin{eqnarray}
    \left(\begin{matrix}
       &                    &                    &    &  t &                    &                    &    & -t &                    &                    &    \\
       & -\frac{2t^2}{U}    &  \frac{2t^2}{U}    &    &    &  t - \frac{t^2}{U} &  \frac{t^2}{U}     &    &    & -\frac{t^2}{U}     &  \frac{t^2}{U} - t &    \\
       &  \frac{2t^2}{U}    & -\frac{2t^2}{U}    &    &    &  \frac{t^2}{U}     &  t - \frac{t^2}{U} &    &    &  \frac{t^2}{U} - t & -\frac{t^2}{U}     &    \\
       &                    &                    &    &    &                    &                    &  t &    &                    &                    & -t \\
     t &                    &                    &    &    &                    &                    &    &  t &                    &                    &    \\
       & -\frac{t^2}{U} + t &  \frac{t^2}{U}     &    &    & -\frac{2t^2}{U}    &  \frac{2t^2}{U}    &    &    &  t - \frac{t^2}{U} &  \frac{t^2}{U}     &    \\
       &  \frac{t^2}{U}     & -\frac{t^2}{U} + t &    &    &  \frac{2t^2}{U}    & -\frac{2t^2}{U}    &    &    &  \frac{t^2}{U}     &  t - \frac{t^2}{U} &    \\
       &                    &                    &  t &    &                    &                    &    &    &                    &                    &  t \\
    -t &                    &                    &    &  t &                    &                    &    &    &                    &                    &    \\
       & -\frac{t^2}{U}     &  \frac{t^2}{U} - t &    &    &  t - \frac{t^2}{U} &  \frac{t^2}{U}     &    &    & -\frac{2t^2}{U}    &  \frac{2t^2}{U}    &    \\
       &  \frac{t^2}{U} - t & -\frac{t^2}{U}     &    &    &  \frac{t^2}{U}     &  t - \frac{t^2}{U} &    &    &  \frac{2t^2}{U}    & -\frac{2t^2}{U}    &    \\
       &                    &                    & -t &    &                    &                    &  t &    &                    &                    &    
\end{matrix}\right).
\label{eq:Effe_matrix1}
\end{eqnarray}
When $i$ and $k$ are next-nearest neighbors, $H_{\text{eff}}$ describes an effective next-nearest-neighbor (NNN) hopping process, represented by the matrix
\begin{eqnarray}
    \left(\begin{matrix}
  &  &  &  & t &  &  &  &  &  &  &  \\
  & -\frac{2t^2}{U} &  \frac{2t^2}{U} &  &  & t &  &  &  & -\frac{t^2}{U} &  \frac{t^2}{U} &  \\
  &  \frac{2t^2}{U} & -\frac{2t^2}{U} &  &  &  & t &  &  &  \frac{t^2}{U} & -\frac{t^2}{U} &  \\
  &  &  &  &  &  &  & t &  &  &  &  \\
t &  &  &  &  &  &  &  & t &  &  &  \\
  & t &  &  &  &  &  &  &  & t &  &  \\
  &  & t &  &  &  &  &  &  &  & t &  \\
  &  &  & t &  &  &  &  &  &  &  & t \\
  &  &  &  & t &  &  &  &  &  &  &  \\
  & -\frac{t^2}{U} &  \frac{t^2}{U} &  &  & t &  &  &  & -\frac{2t^2}{U} &  \frac{2t^2}{U} &  \\
  &  \frac{t^2}{U} & -\frac{t^2}{U} &  &  &  & t &  &  &  \frac{2t^2}{U} & -\frac{2t^2}{U} &  \\
  &  &  &  &  &  &  & t &  &  &  &  
\end{matrix}
    \right).
\end{eqnarray}
To proceed, we adopt a slave-boson representation~\cite{RevModPhys.78.17,PhysRevB.38.5142}. This formalism introduces a slave-boson operator $d^{\dagger}_i$ and a fermionic creation operator $f^{\dagger}_{i,\sigma}$, which act on the vacuum state as $d^{\dagger}_i\left|0\right\rangle=\left|\uparrow,\downarrow\right\rangle_i$ and $f^{\dagger}_{i,\sigma}\left|0\right\rangle=\left|\sigma\right\rangle_i$ ($\sigma=\uparrow, \downarrow$). Assuming a spin-$\uparrow$ polarized bubble without loss of generality, the effective NNN hopping term in the slave-boson representation can be written as
\begin{eqnarray}
\nonumber
    H^{(i, j, k)}_{\text{eff}}&=& \frac{t^2}{U} d^{\dagger}_k f_{k,\uparrow} n_{j,\downarrow} f^{\dagger}_{i,\uparrow}d_{i} +{\rm H. c.} \\
    &\approx& \frac{t^2}{U} \chi^{\uparrow}_{ik}\left(\frac{1}{2}-m\right)d^{\dagger}_k d_i+{\rm H.c.},
\end{eqnarray}
where $m=\left|\left\langle S^z_j \right\rangle\right|$ denotes the magnitude of the background magnetization within the bubble, and $\chi^{\uparrow}_{ik}=\left\langle f_{k,\uparrow} f^{\dagger}_{i,\uparrow}\right\rangle$ represents the spinon hopping parameter, which is omitted for brevity as it merely scales the overall hopping magnitude. 
Finally, the effective low-energy Hamiltonian is given by
\begin{eqnarray}
H_{\text {eff}}\sim \sum_{\langle i ,j \rangle} Zt d^{\dagger}_{i}d_{j}+2\frac{t^2}{U}\left(\frac{1}{2}-m\right)\sum_{\langle \langle i ,k  \rangle \rangle} d^{\dagger}_{i}d_{k} +{\rm H.c.},
\label{effective1}
\end{eqnarray}
where the kinetic renormalization factor $Z$ is introduced into the nearest-neighbor hopping to capture the influence of the polaron. The $\mathcal{O}(t^2/U)$ corrections to the nearest-neighbor hopping in Eq.~(\ref{eq:Effe_matrix1}) are neglected, as they remain small compared to the leading-order scale $t$. The factor of 2 in the NNN term stems from the summation over two distinct intermediate paths on the triangular lattice. In Eq.~(\ref{effective1}), the operator $d$ operates in a singly occupied spin background, behaving as a hard-core boson (or spinless fermion) due to the Pauli exclusion principle.

Following Eq.~(\ref{effective1}), the resulting dispersion relation takes the form
\begin{eqnarray}
\varepsilon({\bold k})=2\tilde{t}\left[2 \cos(\frac{k_x}{2}) \cos(\frac{\sqrt{3}}{2}k_y) + \cos(k_x) \right]+2t' \left[ 2\cos(\frac{3k_x}{2}) \cos(\frac{\sqrt{3}}{2}k_y) + \cos(\sqrt{3}k_y)\right],
\end{eqnarray}
with $\tilde{t}=Zt$ and $t'=\frac{2t^2}{U}(\frac{1}{2}-m)$.
At doping $\delta=0.5$, expanding the dispersion relation in the vicinity of the Van Hove point $\mathbf{M}=(0,\frac{2\pi}{\sqrt{3}})$ up to quartic order yields
\begin{eqnarray}
\varepsilon({\bold M}+ {\bold q})=-\frac{1}{2}(\tilde{t}-9t')q^2_x+\frac{3}{2}(\tilde{t}-t')q^2_y+\frac{1}{96}(7\tilde{t}-81t')q^4_x-\frac{3}{32}(\tilde{t}-7t')q^4_y-\frac{3}{16}(\tilde{t}+9t')q^2_x q^2_y + {\mathcal O}(q^6).
\label{dispersion}
\end{eqnarray} 
A high-order Van Hove singularity (HOVHS) can emerge when the quadratic $\mathcal{O}(\mathbf{q}^2)$ terms in Eq.~(\ref{dispersion}) are strongly suppressed, as the system approaches the critical ratios $\tilde{t}\approx t'$ or $\tilde{t}\approx 9t'$. 
While the former is energetically inaccessible in this model, the latter provides a natural route toward realizing a HOVHS through strong correlations. By employing the effective hopping amplitudes from the effective Hamiltonian, this condition can be explicitly expressed as
\begin{equation}
\frac{Zt}{2\frac{t^2}{U}\left(\frac{1}{2}-m\right)} = 9 \quad \implies \quad Z = \frac{18t}{U}\left(\frac{1}{2}-m\right).
\label{Eq:Z2}
\end{equation}
In the presence of the Nagaoka polaron bubble, the local magnetization satisfies $0 < m < 1/2$. Substituting the typical interaction strength $U/t=12$ into Eq.~(\ref{Eq:Z2}) reveals that the kinetic renormalization factor is bounded by $0 < Z < 3/4$, as discussed in the main text. On the other hand, due to the particle-hole asymmetry of the triangular lattice, hole doping drives the chemical potential away from the $\mathbf{M}$ point, thereby precluding the realization of a high-order Van Hove singularity.

\section{Kinetic Renormalization Factor $Z$ in Interacting Systems}
For a quantum many-body system, the Matsubara Green's function is given by
\begin{eqnarray}
G(\mathbf{k},i\omega_n)=\frac{1}{i\omega_n+\mu-\varepsilon_{\mathbf{k}}-\Sigma(\mathbf{k},i\omega_n)}.
\label{Eq:Z1}
\end{eqnarray}
At finite temperatures, the self-energy is evaluated at discrete Matsubara frequencies. To capture the low-energy effective physics, we analyze its behavior in the low-frequency limit. By treating $\omega_n$ as a continuous variable $\omega$ and exploiting the symmetries of the self-energy—namely, $\mathrm{Re}\Sigma(\mathbf{k}, i\omega)$ being even and $\mathrm{Im}\Sigma(\mathbf{k}, i\omega)$ being odd with respect to $\omega$—the leading-order Taylor expansion can be written as
\begin{eqnarray}
\Sigma(\mathbf{k}, i\omega) \approx \mathrm{Re}\Sigma(\mathbf{k}, 0) + i\omega \left. \frac{\partial \mathrm{Im}\Sigma(\mathbf{k}, i\omega)}{\partial \omega} \right|_{\omega \rightarrow 0}.
\end{eqnarray}
Substituting this expansion back into Eq.~(\ref{Eq:Z1}), we obtain the low-energy asymptotic form of the Green's function:
\begin{eqnarray}
G(\mathbf{k}, i\omega) \approx \frac{1}{i\omega \left[ 1 - \left. \frac{\partial \mathrm{Im}\Sigma(\mathbf{k}, i\omega)}{\partial \omega} \right|_{\omega\rightarrow0} \right] + \mu - \varepsilon_{\mathbf{k}} - \mathrm{Re}\Sigma(\mathbf{k}, 0)}.
\end{eqnarray}
By factoring out the frequency-dependent coefficient in the denominator, we can rewrite the Green's function in a renormalized form,
\begin{eqnarray}
G(\mathbf{k}, i\omega) \approx \frac{Z}{i\omega - Z\left[\varepsilon_{\mathbf{k}} - \mu + \mathrm{Re}\Sigma(\mathbf{k}, 0)\right]}.
\end{eqnarray}
Here, the factor $Z$ emerges from the low-frequency limit of the continuous Matsubara self-energy, defined as
\begin{eqnarray}
Z = \left( 1 - \left. \frac{\partial \mathrm{Im} \Sigma(\mathbf{k}, i\omega)}{\partial \omega} \right|_{\omega \rightarrow 0} \right)^{-1}.
\end{eqnarray}
In our numerical simulations, the zero-frequency derivative $\left.\frac{\partial \mathrm{Im}\Sigma}{ \partial \omega} \right|_{\omega \rightarrow 0}$ is determined by fitting the imaginary part of the self-energy at the lowest several Matsubara frequencies and extrapolating to the $\omega \rightarrow 0$ limit. While $Z$ conventionally denotes the quasiparticle weight in Fermi liquid theory~\cite{PhysRevB.86.085133,10.21468/SciPostPhys.17.3.072}, we here interpret it as a kinetic renormalization factor to account for the breakdown of the quasiparticle picture in the strongly correlated regime.

\section{Density of States near a High-Order Van Hove Singularity}
In this section, we demonstrate the power-law divergence of the density of states (DOS) arising from the HOVHS.
Given the quartic dispersion $\varepsilon(\mathbf{q}) \sim q_x^4 - q_y^2$ in the Nagaoka supermetal regime, the DOS is obtained as
\begin{eqnarray}
\nonumber
\rho(\varepsilon)&=&\int^{\infty}_{-\infty} \int^{\infty}_{-\infty} \frac{dq_x dq_y}{(2\pi)^2}\delta(\varepsilon-q^4_x+q^2_y)\\
&=&\frac{1}{4\pi^2}\int_{q^4_x>\varepsilon}\frac{dq_x}{\sqrt{q^4_x-\varepsilon}}.
\label{Eq:HDOS1}
\end{eqnarray}
When $\varepsilon>0$, by utilizing the even parity of the integrand and applying the substitution $q_x = \varepsilon^{1/4}t$, Eq.~(\ref{Eq:HDOS1}) can be written as
\begin{eqnarray}
\nonumber
    \rho(\varepsilon)&=&\frac{1}{2\pi^2}\int^{\infty}_1 \frac{\varepsilon^{1/4}dt}{\sqrt{\varepsilon(t^4-1)}}\\ \nonumber
    &=&\frac{1}{2\pi^2}\varepsilon^{-1/4}\int^{\infty}_1\frac{dt}{\sqrt{t^4-1}}\\
    &=&\frac{C_+}{8\pi^2}\varepsilon^{-1/4},
    \label{Eq:HDOS2}
\end{eqnarray}
where $C_+ = B\left(\frac{1}{4},\frac{1}{2}\right)$ with $B(P,Q)$ denoting the Beta function. A similar derivation for $\varepsilon<0$ yields
\begin{eqnarray}
    \rho(\varepsilon)=\frac{C_-}{8\pi^2}\left(-\varepsilon\right)^{-1/4},
\end{eqnarray}
with $C_- = B\left(\frac{1}{4},\frac{1}{4}\right)$.
Assuming the chemical potential is pinned at the singularity ($\mu = 0$), the DOS near the HOVHS exhibits a power-law divergence with respect to energy,
\begin{equation}
    \rho(\varepsilon) = C |\varepsilon|^{-0.25},
\label{eq:DOS1}
\end{equation}
with $C$ being a constant. 

At a finite temperature $T$, the effective density of states $N(T)$ is given by the convolution of the DOS and the derivative of the Fermi-Dirac distribution function,
\begin{equation}
    N(T) = \int_{-\infty}^{\infty} \rho(\varepsilon) \left( -\frac{\partial f(\varepsilon)}{\partial \varepsilon} \right) d\varepsilon.
\label{eq:DOS2}
\end{equation}
Evaluating the derivative of the Fermi-Dirac distribution yields
\begin{equation}
    -\frac{\partial f(\varepsilon)}{\partial \varepsilon} = \frac{1}{k_B T} \frac{e^{\varepsilon / k_B T}}{\left( e^{\varepsilon / k_B T} + 1 \right)^2}.
\label{eq:DOS3}
\end{equation}
Substituting Eqs.~(\ref{eq:DOS1}) and (\ref{eq:DOS3}) into Eq.~(\ref{eq:DOS2}), one finds
\begin{equation}
    N(T) = \int_{-\infty}^{\infty} C |\varepsilon|^{-0.25} \frac{1}{k_B T} \frac{e^{\varepsilon / k_B T}}{\left( e^{\varepsilon / k_B T} + 1 \right)^2} d\varepsilon.
\end{equation}
To extract the explicit temperature dependence,  we introduce a dimensionless variable $x = \varepsilon/ k_B T$. In terms of $x$, the energy and its differential are expressed as $\varepsilon = x k_B T$ and $d\varepsilon = k_B T dx$, respectively. Substituting these into the integral, we obtain
\begin{equation}
    N(T) = \int_{-\infty}^{\infty} C |x k_B T|^{-0.25} \frac{1}{k_B T} \frac{e^x}{\left( e^x + 1 \right)^2} (k_B T dx).
\end{equation}
By factoring out the temperature-dependent terms, the expression simplifies to
\begin{equation}
    N(T) = C (k_B T)^{-0.25} \int_{-\infty}^{\infty} |x|^{-0.25} \frac{e^x}{\left( e^x + 1 \right)^2} dx.
\end{equation}
Here, the remaining integral is a dimensionless numerical constant, $A = \int_{-\infty}^{\infty} |x|^{-0.25} \frac{e^x}{(e^x + 1)^2} dx$, which is well-converged and independent of temperature. Consequently, the finite-temperature DOS can be written as
\begin{equation}
    N(T) = C \cdot A \cdot (k_B)^{-0.25} \cdot T^{-0.25},
\end{equation}
which establishes the scaling relation
\begin{equation}
    N(T) \propto T^{-0.25}.
\label{eq:DOS4}
\end{equation}

\end{document}